# Escape rate for the power-law distribution in low-to-intermediate damping

Zhou Yanjun, Du Jiulin

*Department of Physics, School of Science, Tianjin University, Tianjin 300072, China*

**Abstract**：Escape rate in the low-to-intermediate damping connecting the low damping with the intermediate damping is established for the power-law distribution on the basis of flux over population theory. We extend the escape rate in the low damping to the low-to-intermediate damping, and get an expression for the power-law distribution. Then we apply the escape rate for the power-law distribution to the experimental study of the excited-state isomerization, and show a good agreement with the experimental value. The extra current and the improvement of the absorbing boundary condition are discussed.

**Keywords**：Escape rate; low-to-intermediate damping; power-law distributions

## 1. Introduction

In 1940, Kramers proposed a thermal escape of a Brownian particle out of a metastable well [1], and according to the very low and intermediate to high dissipative coupling to the bath, he yielded three explicit formulas of the escape rates in the low damping, intermediate-to-high damping (IHD) and very high damping respectively, all of which has been received great attentions and interests in physics, chemistry, and biology etc [2-3]. In the IHD region, he got an expression of escape rate in the infinite barrier (i.e. the barrier height $E_C \gg k_B T$) and successfully extended it to high damping region; in the low damping region, he derived a rate in energy diffusion regime; as for the intermediate region, he had not given an expression, which was known as Kramers turnover problem. Later, plenty of researches had been continued. Carmeli et al derived an expression for the escape rate in the Kramers model valid for the entire friction coefficient by assuming that the stationary solutions of the low damping and moderate-to-high damping overlap in some region of phase space and are equal to each other (see Eq. (17) in [4] ); Büttiker et al extended the low damping result to the larger range of damping by reconsidering absorbing boundary condition at the barrier and introducing an extra flux $J_{E>E_C}$ (see Eq.(3.11) in [5] ); Pollak et al got a general expression in non-Markov processes (see Eq.(3.33) in [6]); Hänggi et al introduced a simple interpolation formula (see Eq.(6.1) in [7]) for the arbitrary friction coefficient. However, it has been noticed that the above bridging expressions yield results that agree roughly to within $\leq 20\%$ with the numerically precise answers inside the turnover region; in higher dimensions and for the case of memory friction, these interpolation formulas may eventually fail seriously [7]. At the same time, more attentions need to be paid that the systems studied in above theories are all in thermal equilibrium and the distributions all follows a Maxwell-Boltzmann (MB) distribution, $\rho_{eq}(E) = \rho_0 e^{-E/k_B T}$, where $E$ is the energy, $\rho_0$ is the normalization constant, $k_B$ is the Boltzmann constant, and $T$ is the temperature. It should be considered that a complex system far away from equilibrium has not to relax to a thermal equilibrium state with MB distribution, but often asymptotically approaches to a nonequilibrium stationary-state with power-law distributions. In these situations, the Kramers escape rate should be restudied.

In fact, plenty of the theoretical and experimental studies have shown that non-MB distributions or power-law distributions are quite common in some nonequilibrium complex systems, such as in glasses [8,9], disordered media [10-12], folding of proteins [13], single-molecule conformational dynamics [14,15], trapped ion reactions [16], chemical kinetics, and biological and ecological population dynamics [17, 18], reaction–diffusion processes [19], chemical reactions [20], combustion processes [21], gene expression [22], cell reproductions [23], complex cellular networks [24], small organic molecules [25], and astrophysical and space plasmas [26]. The typical forms of such power-law distributions include the noted $\kappa$-distributions in the solar wind and space plasmas [26,27], the $q$-distributions in complex systems within nonextensive statistics [28], and the $\alpha$-distributions noted in physics, chemistry and elsewhere like $P(E) \sim E^{-\alpha}$ with an index $\alpha > 0$ [16,19,20,25,29]. These power-law distributions may lead to processes different from those in the realm governed by Boltzmann-Gibbs statistics with MB distributions. Simultaneously, a class of statistical mechanical theories studying the power-law distributions in complex systems has been constructed, for instance, by generalizing Boltzmann entropy to Tsallis entropy [28], by generalizing Gibbsian theory [30] to a system away from thermal equilibrium, and so forth. Recently, a stochastic dynamical theory of power-law distributions has been developed by means of studying the Brownian motion in a complex system [31,32], which lead the new fluctuation-dissipation relations (FDR) for power-law distributions, a generalized Klein-Kramers equation and a generalized Smoluchowski equation. Based on the statistical theory, one can generalize the transition state theory (TST) to the nonequilibrium systems with power-law distributions [33]; one can study the power-law reaction rate coefficient for an elementary bimolecular reaction [34], the mean first passage time for power-law distributions [35], and the escape rate for power-law distributions in the overdamped systems [36].

In this work, the Kramers escape rate for power-law distributions in the low-to-intermediate damping (LID) will be studied. The paper is organized as follows. In section 2, a generalized escape rate in the LID region is obtained for the power-law distribution and compared with the results of the low damping Kramers' escape rate, and then we apply our theory to the excited-state isomerization of 2-alkenylanthracene in alkane. Further discussion of extra current is given in section 3, and finally the conclusion is made in section 4.

**2. Escape rate for the power-law distribution in the LID**

We have mentioned in the introduction that Büttiker et al. got a Kramer's escape rate in a wider frictional range on the assumption that the system follows the thermal equilibrium distribution. However, for the low damping systems, it is always nonequilibrium. Because the coupling to the bath is very weak and the time to reach thermal equilibrium is very long in low damping systems, the escape of particles may be established before thermal equilibrium, and thus nonequilibrium effects dominate the process [37]. Thereby, the nonequilibrium distribution, such as $\kappa$-distribution, may be used here.

Low damping or small viscosity means that the Brownian forces cause only a tiny perturbation in the undamped energy, so it is helpful to replace the momentum by the energy. In the energy region, the Klein-Kramers equation can be written [3] as

$$\frac{\partial \rho}{\partial t} = \frac{\omega(I)}{2\pi} \frac{\partial}{\partial E}(\gamma I \rho) + \frac{\omega(I)}{2\pi} \frac{\partial}{\partial E}\left(D \frac{\omega(I)}{2\pi} \frac{\partial \rho}{\partial E}\right), \qquad (1)$$

where $\omega$ is the angular frequency of oscillation frequency and it satisfies $\omega(I) = 2\pi dE/dI$, $D$ is the diffusion coefficient, $\gamma$ is the friction coefficient, $I$ is the action defined as $I(E) = \oint_{E=Const} p dx$. In energy space, the continuity equation [3] is

$$\frac{\partial \rho}{\partial t} = -\frac{\omega(I)}{2\pi} \frac{\partial J}{\partial E}. \qquad (2)$$

Take Eq. (2) into Eq. (1) and the current $J$ becomes

$$J = -\frac{D\omega(I)}{2\pi} \exp\left(-\int \frac{2\pi I \gamma}{D\omega(I)} dE\right) \frac{\partial}{\partial E}\left(\rho \exp\int \frac{2\pi I \gamma}{D\omega(I)} dE\right)$$

$$= -\frac{D\omega(I)}{2\pi} \rho_s \frac{\partial(\rho \rho_s^{-1})}{\partial E}. \qquad (3)$$

where $\rho_s$ is the stationary-state distribution $\rho_s = Z^{-1} \exp\left(-\int \frac{2\pi \gamma I}{\omega D} dE\right)$, and $Z$ is the normalization constant. In the previous work, we derived the Kramers' escape rate for the power-law distribution in the low damping, and showed that the stationary-state distribution is the power-law $\kappa$-distribution,

$$\rho_s(E) = Z^{-1}(1 - \kappa \beta E)_+^{1/\kappa}, \qquad (4)$$

if the FDR, $D = \frac{2\pi}{\omega} \gamma I \beta^{-1}(1 - \kappa \beta E)$, is satisfied (see the Appendix in [35]). Thus this FDR is a condition under which the $\kappa$-distribution can be created from the stochastic dynamics of the Langevin equations. When we take the limit of the power-law parameter, $\kappa \to 0$, the power-law $\kappa$-distribution becomes MB distribution, and the FDR becomes the standard one in the traditional statistics.

Supposing that the distribution function $\rho(E)$ can be written as the following form [5], i.e. $\rho(E) = \xi(E)\rho_s(E)$, and Eq. (3) becomes [35],

$$J = -\frac{D\omega(I)}{2\pi}\rho_s(E)\frac{\partial \xi(E)}{\partial E}; \tag{5}$$

integrate over $E$ in both sides of Eq. (5) ($J$ is treated as a constant), we have,

$$J = \left[\xi(E) - \xi(E_C)\right]\left[\int_{k_BT}^{E_C}\frac{2\pi}{D\omega(I)}\rho_s^{-1}(E)dE\right]^{-1}. \tag{6}$$

In the low damping systems, the absorbing boundary conditions are $\xi(E_C) = 0$ and $\xi(E) = 1$ in the bottom and in the well, respectively [35]. Now, in order to extend the low damping regime to the LID regime, the absorbing boundary condition needs to be improved, and an extra current $J_{E>E_C}$ for which the energy $E$ is larger than the barrier energy $E_C$ is considered [5], i.e. $\xi(E_C) \neq 0$. Therefore, the continuity equation is rewritten as

$$\frac{\partial \rho}{\partial t} = -\frac{\omega(I)}{2\pi}\frac{\partial(J + J_{E>E_C})}{\partial E}. \tag{7}$$

When the system reaches the steady state, $J_{E>E_C}$ and $J$ keep balance, Eq.(7) becomes,

$$\frac{\partial J}{\partial E} = -\frac{\partial J_{E>E_C}}{\partial E}, \tag{8}$$

and the left side of Eq.(8) is the current $E < E_C$ induced, so it is still Eq.(6); the right of Eq.(8) comes from the complete integral of the phase-space density multiply velocity at the location of the barrier in the momentum space [5],

$$\begin{aligned}J_{E>E_C} &= \int_{-\infty}^{\infty}\xi(x_C,v)\rho_s(x_C,v)vdp \\ &= \alpha\int_{-\infty}^{\infty}\xi(E)\rho_s(E)dE, \quad E \geq E_C,\end{aligned} \tag{9}$$

where the factor $\alpha$ is a constant of order unit. The reason for introducing $\alpha$ is that in the low damping, the particles move along the orbit of constant energy and the phase-space density increases as one moves them away from the barrier and into the well, due to particles boiling up into this energy range. Thus, the actual distribution function $\rho(x,v)$, at the barrier peak, differs from the average distribution function $\rho(E)$. This is taken into account in Eq. (9) by the factor $\alpha$ [3].

Take the derivative of the energy for Eq. (5), we have

$$\frac{\partial J}{\partial E} = -\frac{\partial}{\partial E}\left[\frac{\omega(I)D\rho_s(E)}{2\pi Z}\frac{\partial \xi(E)}{\partial E}\right]. \tag{10}$$

Eq.(9) and Eq.(10) are then brought together into Eq.(8),

$$\frac{\partial}{\partial E}\left[\frac{\omega(I)D\rho_s}{2\pi}\frac{\partial \xi(E)}{\partial E}\right] = \alpha\xi(E)\rho_s(E). \tag{11}$$

We use the generalized FDR to go on the further simplification. Because a relatively narrow energy range above $E=E_C$ is concerned, $I$ can be taken as sensibly constant, i.e. $I=I_C$ [5], the frequency, i.e. $\omega(I)/2\pi \approx \omega_0/2\pi$. In the low damping, the friction coefficient has less effect on the energy, so it also can be taken as a constant, i.e. $\gamma = \gamma_C$. Therefore, Eq. (11) becomes,

$$\frac{\omega_0 D(E)}{2\pi}\xi'' + \frac{\omega_0}{2\pi}\left(\frac{\partial D(E)}{\partial E} - \frac{2\pi\gamma_C I_C}{\omega_0}\right)\xi' - \alpha\xi = 0, \tag{12}$$

within a small energy range above $E_C$ one can assume essentially a constant diffusion coefficient [38], we might as well take

$$D\big|_{E\approx E_C} = \frac{2\pi}{\omega_0}\gamma_C I_C \beta^{-1}(1-\kappa\beta E_C), \quad \frac{\partial D(E)}{\partial E}\bigg|_{E\approx E_C} = 0, \tag{13}$$

hence Eq. (12) turns into a conventional ordinary differential equation for $E$,

$$\gamma_C I_C \beta^{-1}(1-\kappa\beta E_C)\xi'' - \gamma_C I_C \xi' - \alpha\xi = 0, \tag{14}$$

which has a solution,

$$\begin{aligned}\xi(E) = &C_1 \exp\left\{\frac{\beta E}{2(1-\kappa\beta E_C)}\left[1+\sqrt{1+\frac{4\alpha(1-\kappa\beta E_C)}{\gamma_C I_C \beta}}\right]\right\} \\ &+ C_2 \exp\left\{\frac{\beta E}{2(1-\kappa\beta E_C)}\left[1-\sqrt{1+\frac{4\alpha(1-\kappa\beta E_C)}{\gamma_C I_C \beta}}\right]\right\}\end{aligned}, \tag{15}$$

where $C_1$ and $C_2$ are two integral constants respectively. The density should be the definite value when the energy increases, thus the first term of the right side of Eq.(15) is abandoned and the solution is written as,

$$\xi(E) = C_2 \exp\left\{\frac{\beta E}{2(1-\kappa\beta E_C)}\left[1-\sqrt{1+\frac{4\alpha(1-\kappa\beta E_C)}{\gamma_C I_C \beta}}\right]\right\}. \tag{16}$$

Make a substitution,

$$s = \frac{1}{2}(1-\kappa\beta E_C)^{-1}\left[1-\sqrt{1+\frac{4\alpha(1-\kappa\beta E_C)}{\gamma_C I_C \beta}}\right], \tag{17}$$

and Eq.(16) becomes a more convenient form,

$$\xi(E) = C_2 e^{s\beta E}. \tag{18}$$

In order to keep $\xi(E)$ continuous at $E=E_C$, assuming Eq.(18) has the following form [5],

$$\xi(E) = \xi(E_C) e^{s\beta(E-E_C)}. \tag{19}$$

Next we derive the expression of $\xi(E_C)$ according to equating the current of $E < E_C$ and the current of $E > E_C$ at $E = E_C$. The current of $E > E_C$ is

$$\begin{aligned} J &= -\frac{\omega_0 D \rho_s(E)}{2\pi}\frac{\partial \xi(E)}{\partial E} \\ &= -\frac{s\xi(E_C)\gamma_C I_C}{Z}(1-\kappa\beta E_C)(1-\kappa\beta E)^{1/\kappa} e^{s\beta(E-E_C)}, \end{aligned} \tag{20}$$

and the current of $E<E_C$ (Ref. [35] Appendix (A.22) where $E_b$ is replaced by $E_C$) is

$$\begin{aligned} J &= \frac{1-\xi(E_C)}{\int_{k_B T}^{E_C}\frac{2\pi}{D\omega(I)}\rho_s^{-1}dE} \\ &= \frac{\gamma_C I_C (1-\kappa)}{Z}\left[1-\xi(E_C)\right](1-\kappa\beta E_C)^{\frac{1}{\kappa}}. \end{aligned} \tag{21}$$

Equating the above two expressions at $E=E_C$ gives

$$\xi(E_C) = \frac{1-\kappa(1+s\beta E_C)}{1-(\kappa+s)}. \tag{22}$$

One of the most significant escape rate theories is the flux over population theory [2,3,7]. If the steady-state current $J$ and the (nonequilibrium) population inside the initial domain $n$ are got, the rate of escape is then given by the ratio, $k = J/n$. Hence, according to the appendix Eq.(A.9) in Ref.[35], the escape rate for the power-law distribution can be finally obtained as

$$k_\kappa = \frac{\gamma_C I_C \beta (1-\kappa)(1+\kappa)\left[\sqrt{1+\frac{4\alpha(1-\kappa\beta E_C)}{I_C \gamma_C \beta}} - 1\right]}{\sqrt{1+\frac{4\alpha(1-\kappa\beta E_C)}{I_C \gamma_C \beta}} - 1 + 2(1-\kappa)(1-\kappa\beta E_C)} k_{\kappa-\mathrm{TST}}. \tag{23}$$

where $k_{\kappa-\mathrm{TST}}$ is the TST rate for the power-law distribution if we choose $V_a=0$ and $\kappa=\nu-1$ (see Eq. (63) and Eq. (64) in [33] ). In the limit $\kappa \to 0$, Eq. (23) can return to the traditional result of the escape rate for MB distribution (see Eq.(3.11) in [5]),

$$k_{\mathrm{BHL}} = \frac{\sqrt{1+\frac{4\alpha}{I_C \gamma_C \beta}} - 1}{\sqrt{1+\frac{4\alpha}{I_C \gamma_C \beta}} + 1} \frac{\omega_0 \gamma_C I_C \beta}{2\pi} e^{-\beta E_C}. \tag{24}$$

As the friction coefficient tends to zero $\gamma_C \to 0$, Eq.(23) reduces to

$$k_\kappa = \frac{\gamma_C I_C \omega_0 \beta (1-\kappa)}{2\pi} \chi_\kappa (1-\kappa\beta E_C)^{\frac{1}{\kappa}}, \tag{25}$$

with

$$\chi_\kappa = \begin{cases} -\kappa \Gamma^2\left(-\frac{1}{\kappa}\right) \Big/ \Gamma^2\left(-\frac{1}{\kappa}-\frac{1}{2}\right), & (-2<\kappa<0) \\ \kappa \Gamma^2\left(\frac{1}{\kappa}+\frac{3}{2}\right) \Big/ \Gamma^2\left(\frac{1}{\kappa}+1\right), & (\kappa>0) \end{cases}$$

which coincides with the Kramers escape rate in the low damping for the power-law distribution (see Appendix (A.14) in [35]); therefore Eq.(23) contains the low damping region. When friction coefficient tends to infinity $\gamma_C \to \infty$, Eq.(23) becomes

$$k_\kappa = \frac{\alpha \omega_0}{2\pi} \chi_\kappa (1-\kappa\beta E_C)^{\frac{1}{\kappa}} = \alpha(\kappa+1) k_{\kappa-\mathrm{TST}}. \tag{26}$$

In Fig.1, we plot the low damping, LID and IHD rates for the power-law distribution, which are all normalized by the TST rate for MB distribution. The IHD rate for the power-law distribution is derived in the Appendix (see Eq.(A.15)). We see from Fig.1 that two solid curves of the LID with MB distribution and power-law distribution almost overlap with the ones of the low damping, which has been explained in the above deduction. As the damping $\gamma$ increases, the traditional curves of the LID and IHD intersect at about $\gamma=0.8$ and the ratio of the rate normalized by the TST result is about $k_{\mathrm{BHL}}/k_{\mathrm{TST}} = 0.93$; whereas they intersect at about $\gamma=5$ for the power-law distribution and the ratio is about $k_\kappa/k_{TST}=0.64$. When the larger power-law parameter $\kappa$ is taken, the intersection point will move to the higher damping. With much higher damping $\gamma$, two solid curves both approach to the TST rate deviating from each curve of IHD. Though the result of the LID overestimates the rate in the very higher damping, the transition from the low damping to LID is reasonably achieved and predicts a lower escape rate than the Kramers results.

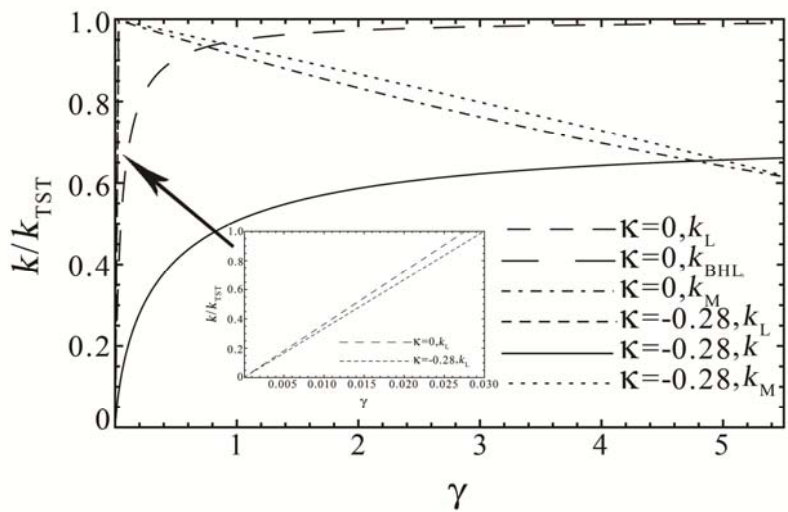

Fig.1. Theoretical estimation of the escape rate normalized by the TST result in three damping ranges for power-law parameter $\kappa=-0.28$, the barrier height $E_C=5.77k_BT$, the frequency of the barrier $\omega_C=5.45$, the mass of particles $m=1$, $\beta=1$ and $\alpha=1$ [5, 38]. $k_L$ and $k_M$ are the Kramers' low damping rate and IHD rate respectively. The inset showed the enlarged parts for the rate in the low damping.

Then we apply our result to the experiment and check the prediction. Hara et al studied the Kramers turnover behavior for the excited-state isomerization of 2-alkenylanthracene in alkane at the high pressure [39]. The experimental material was 2-(2-propenyl) anthracene (22PA), synthesized using the method of Stolka et al. [40] and purified by TLC. Steady-state and time-resolved fluorescence spectra in supercritical (SC) ethane (99.95%) and SC $CO_2$ (99.999%) were measured at 323 K and at pressures up to 15.1 and 17.4 MPa respectively. FIG.3 of [39] indicated a clear demonstration of the Kramers turnover behavior with increasing the viscosity. At the same time, the interaction (i.e. dynamic solvent effect) between the solute and solvent was also studied (see FIG.5 in [39]), and the consequence can be well explained by our LID result. These parameters we adopt in Fig.1 keep the same with the experimental data in [39], i.e. activation energy, $E_0=5.77k_BT$, the mass of particle, $3.223\times10^{-25}kg$, and the barrier top frequency, $\sqrt{m}\omega_b = 5.45 = \omega_C$. At the turning point, our result $k_\kappa/k_{TST}=0.64$ with the power-law parameter $\kappa=-0.28$ agrees with the experimental value $\kappa_{max}(=k_f/k_{TST})=0.64$ (see Table II in [39]). It is therefore concluded that our theory represents excellently the experimental result as compared to the traditional theory.

## 3. Further discussion of extra current

In Section 2, an extra current $J_{E>E_C}$ is introduced. The problem naturally arises whether it exists or not and how the friction coefficient affects it. Now we make further discussion. First we calculate the probability $P(E)dE$ that an escaping particle has an energy $E$ between $E$ and $E+dE$. The current $J$ that $E<E_C$ produces is zero due to the steady-state distribution; the excess energy $\Delta E=E-E_C$ is introduced and the probability is

$$P(E)dE = \frac{dJ_{E>E_C}}{J+J_{E>E_C}} = \frac{\alpha\rho(E)dE}{\int_{E_C}^{\infty}\alpha\rho(E)dE} \qquad (27)$$

$$= \frac{e^{s\beta(E-E_C)}(1-\kappa\beta E)^{1/\kappa}dE}{(1-\kappa\beta E_C)^{1/\kappa}\int_0^{\infty}e^{s\beta\Delta E}\left(1-\frac{\kappa\beta\Delta E}{1-\kappa\beta E_C}\right)^{1/\kappa}d\Delta E}.$$

The average of the excess energy is then given by

$$\langle\Delta E\rangle = \int_{E_C}^{\infty}(\Delta E)P(E)dE = \int_0^{\infty}d(\Delta E)\frac{\Delta E\, e^{s\beta\Delta E}\left(1-\frac{\kappa\beta\Delta E}{1-\kappa\beta E_C}\right)^{1/\kappa}}{\int_0^{\infty}e^{s\beta\Delta E}\left(1-\frac{\kappa\beta\Delta E}{1-\kappa\beta E_C}\right)^{1/\kappa}d(\Delta E)}. \qquad (28)$$

We do numerical integral about Eq. (28) and plot the average energy in the extremely low damping and the extremely high damping, respectively, for different power-law parameters; other parameters are taken as $\alpha=1$ [5, 38] and $\beta E_C=10$.

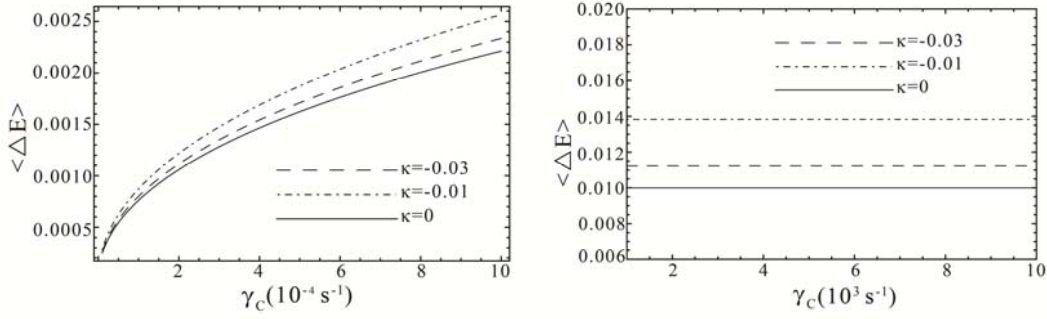

(*a*) Extremely low damping, $\gamma_C \to 0$      (*b*) Extremely high damping, $\gamma_C \to \infty$

Fig.2 The influence of the friction coefficient on the average energy for different power-law parameters

In Fig.2, with decreasing damping the average energy of escaping particles decreases to zero both for the MB distribution and the power-law $\kappa$-distribution. So there does not exist the extra current $J_{E>E_C}$ in extremely low damping, i.e. $\rho(E_C)=0$ which Kramers had ever assumed, and the Kramers low damping rate corresponds to the underdamped case. For extremely high damping, the average energy approaches the constant independent of the friction coefficient in both cases, and thus there definitely exists the extra current $J_{E>E_C}$. Thereby, we get a conclusion that when the damping is extremely low, the absorbing boundary condition at the barrier which was always used in the past is right; once the damping is not very low, the absorbing boundary condition becomes an approximation and then needs to be improved by taking the extra current $J_{E>E_C}$ into account.

## 4. Conclusion

Many physical, chemical and biological systems are complex, and usually open and nonequilibrium. In fact, a complex system far away from equilibrium does not have to relax to thermal equilibrium with a MB distribution, but often asymptotically approaches a stationary nonequilibrium with a power-law distribution. Therefore, the escape rate theory should be reestablished under the framework of the statistics of power-law distributions. According to the flux over population theory, we have extended the result in the low damping to a wider range of the friction coefficient by improving the absorbing boundary condition, and get the expression of escape rate in the low-to-intermediate damping (LID) for the power-law $\kappa$-distribution. When the damping is extremely low, it returns to the Kramers escape rate in the low damping; when the damping is extremely high, it reduces to the TST rate.

We have applied our theory to the experimental study of the excited-state isomerization of 2-alkenylanthracene in alkane, checked the prediction and concluded that the result was a good agreement with the experimental value. Furthermore, we have made the numerical analyses and further discussions about the extra current.

**Appendix**

Particles move in the IHD systems and the process is governed by the Klein-Kramers equation [3],

$$\frac{\partial \rho}{\partial t} = -\frac{p}{m}\frac{\partial \rho}{\partial x} + \frac{\partial}{\partial p}\left(\frac{dV(x)}{dx} + \gamma(x,p)\rho\right) + \frac{\partial}{\partial p}\left(D(x,p)\frac{\partial \rho}{\partial p}\right). \tag{A1}$$

In Eq.(A1), if the coefficients $D(x,p)$ and $\gamma(x,p)$ satisfy the generalized fluctuation-dissipation relation given [31] by

$$D = m\gamma\beta^{-1}(1-\kappa\beta E), \tag{A2}$$

then the stationary-state solution is the power-law $\kappa$-distribution,

$$\rho_s(E) = Z^{-1}(1-\kappa\beta E)_+^{1/\kappa}, \tag{A3}$$

for the energy $E$. In the limit $\kappa \to 0$, the distribution returns to the MB distribution.

Supposing the barrier is located at $x_C$ and the potential can be expanded as a Taylor series about $x_C$, one can write taking the barrier top as the zero of the potential [3],

$$V_1 = -\frac{1}{2}\omega_C^2(x-x_C)^2. \tag{A4}$$

Near the bottom of the well, $x_A$ ($x_A \approx 0$), the potential is approximated by

$$V_2 = -\Delta V + \omega_A^2 x^2/2, \tag{A5}$$

where $\Delta V = V(x_C) - V(x_A)$. Take Eq.(A4) into Eq.(A1), and Eq.(A1) becomes

$$\omega_C^2 x' \frac{\partial \rho}{\partial p} + p \frac{\partial}{\partial x'} \rho - \frac{\partial}{\partial p}\left(\gamma p \rho + D \frac{\partial \rho}{\partial p}\right) = 0, \tag{A6}$$

where $x' \equiv x - x_C$. Now make the substitution [3] and take the power-law steady-state solution at the barrier,

$$\rho \equiv \xi(x', p) \rho_s = \xi(x', p)\left[1 - \kappa \beta \left(p^2 - \omega_C^2 x'^2\right)/2\right]^{1/\kappa}, \tag{A7}$$

substituted into Eq.(A6), combining Eq.(A2) to simplify, we have

$$\omega_C^2 x' \frac{\partial \xi}{\partial p} + p \frac{\partial \xi}{\partial x'} + (\gamma p - \frac{\partial D}{\partial p}) \frac{\partial \xi}{\partial p} - D \frac{\partial^2 \xi}{\partial p^2} = 0. \tag{A8}$$

Here we adopted a special case, i.e. assume the friction coefficient is a constant, $\gamma = \gamma_C$, but the diffusion coefficient is a function of the energy, $\partial D/\partial p = -\kappa \gamma_C p$, then Eq.(A8) becomes

$$\left\{-\omega_C^2 x' + [a - (\kappa+1)\gamma_C] p\right\} \xi' + D \xi'' = 0. \tag{A9}$$

To solve this equation, one wishes to write the coefficients $\xi'$ and $\xi''$ in terms of the single variable, $u \equiv p - ax'$, rather than $x'$ and $p$ [3], where $a$ is an undetermined constant. It can be achieved in a very neat way if one writes

$$-\omega_C^2 x' + \left(a - (\kappa+1)\gamma_C\right) p = \left[a - (\kappa+1)\gamma_C\right] u, \tag{A10}$$

which imposes on $a$ the condition:

$$a_\pm = \frac{1}{2}(\kappa+1)\gamma_C \pm \frac{1}{2}\sqrt{(\kappa+1)^2 \gamma_C^2 + 4\omega_C^2}. \tag{A11}$$

Eq.(A9) then takes the form of a conventional ordinary differential equation in $u$,

$$\xi = C \int_u e^{-\int \frac{[a-(\kappa+1)\gamma_C]u'}{D(u')} du'} du', \tag{A12}$$

where $C$ is an integral constant and $a$ takes $a_+$ so as to make the distribution finite.

The probability current $J$ crossing the barrier can be obtained by integrating for $p\rho$ over $p$ from minus infinity to infinity,

$$J = \int_{-\infty}^{\infty} p\rho dp = C\int_{-\infty}^{\infty} p\left(1-\kappa\beta\frac{p^2}{2}\right)^{1/\kappa} dp \int_{-\infty}^{p} dp' e^{-\int_{p'}\frac{[a-(\kappa+1)\gamma_C]x}{\beta^{-1}\gamma_C(1-\kappa\beta x^2)}dx} . \quad \text{(A13)}$$

While the number of particles $n$ trapped near the minimum $A$ is

$$n = C\frac{2\pi\chi_\kappa(1+\kappa\beta\Delta V)^{\frac{1}{\kappa}}}{\kappa\beta\omega_A} \int_{-\infty}^{\infty} dp\, e^{-\int_{p'}dp'\frac{[a-(\kappa+1)\gamma_C]p'}{D(p')}} . \quad \text{(A14)}$$

The probability of the escape is therefore the number crossing the saddle line in unit time divided by the number in the well,

$$k = \frac{J}{n} = \frac{\beta(\kappa+1)\int_{-\infty}^{\infty} p\left(1-\kappa\beta\frac{p^2}{2}\right)^{1/\kappa} dp \int_{-\infty}^{p} e^{-\int_{p'}dx\frac{[a-(\kappa+1)\gamma_C]x}{\beta^{-1}\gamma_C(1-\kappa\beta x^2/2)}} dp'}{\int_{-\infty}^{\infty} e^{-\int_{p}dp'\frac{[a-(\kappa+1)\gamma_C]p'}{D(p')}} dp} k_{\kappa-\text{TST}} . \quad \text{(A15)}$$

**Acknowledgments**

This work is supported by the National Natural Science Foundation of China under Grant No 11175128 and by the Higher School Specialized Research Fund for Doctoral Program under Grant No 20110032110058.